\documentclass[12pt,aps]{revtex4}
\usepackage[cp1251]{inputenc}

\input{epsf}
\begin{document}

\title{Slow Light in a Bragg Waveguide}
\author{G.G.Kozlov, V.S.Zapasskii  and V.V.Ovsyankin}

 \hskip20pt
\vskip20pt
\hskip100pt {\it e}-mail:  gkozlov@photonics.phys.spbu.ru
\vskip20pt
\begin{abstract}
One-dimensional defect photonic crystal (Bragg waveguide) is studied from the
viewpoint of the slow light problem. The calculations are presented showing that
in the TiO$_2$/SiO$_2$-based Bragg waveguide one can obtain the group index of $\sim$ 1000
and spatial decay length of $\sim$ 3 mm for a nanosecond-scale pulse. Distortion of
the pulse due to the group index dispersion proves to be acceptable for the
relative pulse delay not exceeding 10. We also analyze propagation of the light
pulse in the Bragg waveguide with a quantum well inside and provide arguments
showing possibility of reaching the group index of $\sim$ 10000.
To the best of our knowledge, analysis of pulse propagation
in a Bragg waveguide in connection with the slow light problem has
 not been performed so far. We will still much appreciate any information about such studies, if any.
\end{abstract}
\maketitle

\section{Introduction and basic results}

In the last decade, a new direction of optics - so-called "slow light" - has
been developed. The goal of research performed in the framework of this trend is
to reduce the light group velocity as much as possible. Importance of these
studies is related to practical need of creating compact delay lines for optical
systems of information processing. At present, a great number of papers are
published devoted to slow light (see, e.g., review \cite{Slepov}). In the most famous of
them \cite{Hau}, the light propagation with a group velocity of 17 m/s has been
reported. However, the light speed reduction, in this experiment, was observed
in a rather narrow spectral range in a medium of supercooled atomic gas. For the
system of information processing, such a huge reduction of the light pulse group
velocity is not required, while the operating bandwidth (inverse pulsewidth)
should be preferentially much broader, e.g., to lie in the range of 10$^9$ Hz. It
is also desirable not to deal with such exodic media as atomic gas cooled below
the point of Bose-Einsten condensation \cite{Hau}. In the light of these remarks, it
seems promising to study slow light in photonic crystals and
photonic-crystal-based waveguides \cite{Krauss,Toshihiko}. The law of dispersion of electromagnetic
waves $q(\omega)$ (here, $q$ and $\omega$ are, respectively, the wavenumber and frequency of the
wave) in these systems is characterized by the presence of frequencies where
$dq/d\omega$ tends to infinity and the group velocity of light tends to zero. In
photonic crystals, these frequencies correspond to edges of the allowed
electromagnetic zones. Similar features of the dispersion law are displayed by
the waveguides formed by an extended defect in a photonic crystal, along which
an electromagnetic wave localized around this defect may propagate \cite{Notomi}. The
papers devoted to this version of slow light consider, as a rule, propagation of
a wave along a linear defect in a planar photonic crystal. Such a defect
photonic crystal represents a dielectric plate with regularly arranged apertures
(usually in the form of a hexagonal lattice), with the linear defect being a
break of this regularity along some line (e.g., a skip of one line of the
apertures in the lattice). Mathematical analysis of the electromagnetic wave
localized around such a defect is hampered because even the problem of finding
forbidden electromagnetic zones in a defectless planar photonic crystal, needed
tor this analysis, proves to be rather complicated.

The simplest problem of this
type, which still may be important for applications, is the one about
propagation of electromagnetic wave along a defect in a one-dimensional photonic
crystal. Such a system comprises two closely spaced Bragg mirrors forming the
well-known Bragg resonator (or Bragg waveguide). In spite of simplicity of such
a system, the slow-light propagation in such a system, to the best of our
knowkedge, has not been so far analyzed. This analysis is the objective of the
present paper. For the mentioned simplest system, the problem of propagation of
an electromagnetic pulse in a gap between two parallel Bragg mirrors can be
solved exactly. A complete analysis also proves to be possible for the waveguide
comprised of finite Bragg mirrors. In this case, propagation of light is
accompanied by attenuation (leakage), which can be also consistently calculated.
The main result of this paper is calculation of the Bragg waveguide dispersion
law with account for the leakage and demonstration of possibility to delay, in
such a waveguide, the light pulse with a duration of 10$^{-9}$ s by the time
substantially exceeding its duration (by a factor of 10 - 30). In this case, the
value of the group refractive index appears to be 1000 or higher. The
calculations show that fabrication of the Bragg waveguide with acceptable
attenuation (leakage) is practically feasible. In particular, to obtain the
decay length of 2 - 3 mm in the SiO$_2$/TiO$_2$-based Bragg waveguide, it suffices
to apply 15-period Bragg mirrors. To obtain the same attenuation in the
GaAs/AlGaAs-based Bragg waveguide, the number of periods should be close to 100.
To considerably slow down the light pulse in such systems, the pulse carrier
frequency should be chosen to be close to that of the waveguide cutoff $\bar\omega$. At
this frequency the dispersion law of the waveguide shows a singularity where
$dq/d\omega=\infty$. As a further development of the slow-light technique described above, we
analyze the Bragg waveguide with a single quantum well grown inside it. Now, if
the resonance frequency $\omega_0$ of the quantum well lies below the cutoff frequency $\bar\omega$ of
the "empty" waveguide, then the transmission spectrum of such a composite system
shows a window in the frequency region close to $\omega_0$. The dispersion law within
this window appears to be fairly steep which promises further group velocity
reduction.

\section{Bragg waveguide}

In this section, we calculate propagation of electromagnetic wave with the
frequency $\omega$ and wavenumber $q$ in the Bragg waveguide with the structure shown in
Fig. 1. The waveguide consists of two identical Bragg mirrors arranged
symmetrically with respect to the x-y plane at a distance of $l_0$ from this
plane. We analyze the case of the TE-wave, with only y-component of the electric
field being nonzero. Let us number the layers of the structure in the half-space
$z > 0$ as shown in Fig. 1. Let $z_i$ be z-coordinate of the boundary between the
$(i-1)$-th and $i$-th layers of the structure, $l_i =\equiv z_{i+1}-z_i$  is the
  thickness of the $i$-th
layer, and $\varepsilon_0$ is its permittivity. In accordance with the problem formulated
above, we have to find solution of the Maxwell equations for the layer structure
of Fig. 1 with the dependence on time $t$ and $x$-coordinate as  $\sim\exp\imath[\omega t+qx]$.
  In what
follows, for brevity, we do not specify this dependence explicitly,
incorporating it into the field amplitudes
 The Bragg waveguide under consideration has much in common with that comprised
of two metal mirrors, whose transmission spectrum, as is known, is
bound from below by the cutoff frequency. As we will see below,
this property is also characteristic of the Bragg waveguide under
study. When the frequency $\omega$ of the electromagnetic
oscillations in the waveguide exceeds the cutoff frequency
$\bar\omega$, the wavenumber $q$ appears to be a definite real
function of the frequency $\omega$. This function is commonly
referred to as the dispersion law. At the same time, the motion of
electromagnetic field in the waveguide can be specified in terms
of the types of the field oscillations (modes). For this reason,
one waveguide may exhibit many dispersion laws
$q(\omega)=q_m(\omega)$, where $m$  --  is the mode number.

Let us represent the field in the $i$-th layer of the structure (Fig. 1) in the
form of a superposition of two waves with the same $x$-components  $q_x\equiv q$ and opposite
z-components  $\pm q_z$ of the wave vector. Since the field in this layed should satisfy
the Maxwell equations, we have $q^2+q_z^2=\varepsilon_i k^2$. Then the electric field in the $i$-th
 layer will be given by:
\begin{equation}
\hbox{  field $i$-th layer for $z>0$:} \hskip2mm E_y=
A_i\exp\imath\sqrt{\varepsilon_i k^2-q^2}z+
B_i\exp-\imath\sqrt{\varepsilon_i k^2-q^2}z
\end{equation}

Assume that no light is incident on the Bragg mirror from the upper half-space.
Then, there is only outcoming wave in the region $z>z_{n+1}$  i.e., $B_{n+1}=0$. The amplitude
$A_0$ of the wave incident from below onto the upper Bragg mirror and the amplitude
$B_0$ of the wave reflected from it are proportional to each other:
\begin{equation}
  B_0=\ae A_0,
\end{equation}

 with the proportionality factor $\ae=\ae(\omega,q)$ having the sense of the Bragg mirror
reflectivity. Note that $z$-coordinate of the edge of the Bragg mirror is $l_0$
(rather than zero), and the reflectivity $\ae$ should contain the appropriate
phase factor. Let us turn now to the field distribution at $z < 0$. We will number
the layers in this region of the structure in a symmetric way, and will supply
the field amplitudes of the lower Bragg structure with hats:

\begin{equation}
\hbox{  field in the $i$-th layer for $z<0$:} \hskip2mm E_y=
\tilde{A}_i\exp\imath\sqrt{\varepsilon_i k^2-q^2}z+
\tilde{B}_i\exp-\imath\sqrt{\varepsilon_i k^2-q^2}z
\end{equation}
Since the lower Bragg structure is just the upper one reflected in the $x-y$ plane, we should
have:
\begin{equation}
  \tilde{A}_0=\ae \tilde{B}_0.
\end{equation}
 If the Bragg waveguide (Fig.1) contains a quantum well in the plane $z=0$, then the field
amplitudes $A_0, B_0, \tilde{A}_0, \tilde{B}_0$ are connected by the proper
 boundary conditions. If the waveguide
is empty, then $A_0=\tilde{A}_0$ and  $B_0=\tilde{B}_0$, hence:

\begin{equation}
\ae^2(\omega,q)=1
\end{equation}
This equation yields in the implicit way the dispersion law $q(\omega)$ of the Bragg
waveguide. Let us dwell upon some qualitative features of this law. Consider an
ideal Bragg waveguide with infinite number of periods in the mirrors. Then, in
conformity with known properties of the Bragg reflector, one can indicate the
frequency intervals (forbidden bands of the one-dimensional photonic crystal)
within which the reflectivity module equals unity, i.e., the windows of total
reflection. Therefore, for the frequencies matching the total reflection windows,
we have:
\begin{equation}
\ae=\exp\imath\xi
\end{equation}
 With $\xi=\xi(\omega,q)$ being a real function.  Then, Eq. (5) yields:
\begin{equation}
 \xi(\omega,q)=\pi m, \hskip10mm\hbox{where }\hskip2mm m \hskip2mm\hbox{ -- integer}
 \end{equation}

Real solutions of this equation  $q_m(\omega)$ provide the laws of wave dispersion, with the
integer parameter $m$ corresponding to the field oscillation mode number. For the
ideal Bragg waveguide, such solutions are possible only at the frequencies $\omega$
exceeding the cutoff frequency $\bar\omega$, in complete similarity with the waveguide
comprised of two metal mirrors. To get some idea about the form of the
dispersion law of the Bragg waveguide, we assume that spectral positions of the
windows of total reflection and distances between these mirrors are chosen to
form a classical Fabry-Perot interferometer. This will be the case when, e.g.,
the distance between the Bragg mirrors is equal to half the wavelength
corresponding to the central frequency of the reflectivity window. In this case,
Eq. (5) has a solution at $q = 0$, i.e.,$\ae^2(\bar\omega,0)=1$, where $\bar\omega$
 is the resonance frequency of
the Fabry-Perot interferometer at normal incidence. Calculation of the
reflectivity $\ae(\omega,q)$ shows that, in fact, it depends on the
 quantity $s\equiv q^2$. Hence, it
follows that the dispersion law (5) can be rewritten in the form:
\begin{equation}
\Phi(\omega,q^2)\equiv\Phi(\omega,s)=0,\hskip10mm s\equiv q^2
\end{equation}
Where $\Phi(\omega,q^2)=1-\ae^2(\omega,q)$.  In conformity with the aforesaid, we have
  \begin{equation}
  \Phi(\bar\omega,0)=0
  \end{equation}
The dispersion law of the Bragg waveguide at small $q$ and frequencies close to
$\bar\omega$ can be obtained by expanding (8) in series:
\begin{equation}
{\partial\Phi\over\partial\omega}\bigg|_{\omega=\bar\omega, \hskip1mm s=0}
(\omega-\bar\omega)+
{\partial\Phi\over\partial s}\bigg|_{\omega=\bar\omega, \hskip1mm s=0} q^2=0
\end{equation}
  Whence
\begin{equation}
q=G\sqrt{\omega-\bar\omega},\hskip5mm G\equiv\sqrt{-{\partial\Phi/\partial\omega\over \partial\Phi/\partial
s}}\bigg|_{\omega=\bar\omega, \hskip2mms=0}
\end{equation}

Using this formula, one can see that the group refractive index $n_g$ defined as
  \begin{equation}
  n_g=c\hskip2mm{dq\over d\omega}=
 {c\hskip1mmG\over 2\sqrt{\omega-\bar\omega}}=
-c\hskip2mm{\partial\Phi/\partial\omega\over \partial\Phi/\partial
  s}\hskip2mm{1\over 2 q}
  \end{equation}
 turns into infinity at $\omega=\bar\omega$ (at $q=0$),
   which motivates the interest to the waveguide
under study from the viewpoint of obtaining of slow light.

Turn now to the case of a real Brag waveguide with $n<\infty$, for which  $|\ae|<1$ even
for the frequencies matching the windows of total reflection. In this case,
solutions of the dispersion equation (5) $q_m(\omega)$ become complex even in the
spectral regions where they were real for the ideal Bragg waveguide. The
presence of imaginary part in the wavenumber $q_m(\omega)$  corresponds to attenuation of the
wave, with the relevant modes being referred to as leaking. In what follows we
present a consistent mathematical derivation of the expression for reflectivity
  $\ae(\omega,q)$ of the Bragg mirror with a given number of the periods $n$. Though this
derivation is standard \cite{Koz}, the obtained formulas are aimed
at derivation of the complex-valued dispersion laws for waveguides
with leakage. For this reason, it seems sensible to present this
derivation here.

\section{Reflectivity of the Bragg mirror}

To obtain expression for the reflectivity $\ae$, one has to calculate field
distribution in the Bragg structure (Fig. 1). Using the boundary conditions

\begin{equation}
E_y(z_{i+1}-0)=E_y(z_{i+1}+0)
\end{equation}
$$
{\partial E_y\over\partial z}(z_{i+1}-0)={\partial E_y\over\partial z}(z_{i+1}+0)
$$
 The field amplitudes in the $i$-th and $(i+1)$-th layers can be be interconnected by
a linear transformation, which can be written as a product of three transfer
matrices:
\begin{equation}
\left(\matrix{A_{i+1}\cr B_{i+1}}\right)=U_{i+1}^{-1}T_{i\rightarrow i+1}
U_i \left(\matrix{A_i\cr B_i}\right)
\end{equation}
where
\begin{equation}
U_i=\left(\matrix{e^{\imath z_i\sqrt{\varepsilon_i k^2-q^2}} & 0\cr
0 & e^{-\imath z_i\sqrt{\varepsilon_i k^2-q^2}}}\right)
\end{equation}
\begin{equation}
T_{i\rightarrow i+1}={1\over 2}\left(\matrix{
\bigg(1+\sqrt{\varepsilon_i k^2-q^2\over\varepsilon_{i+1} k^2-q^2}\bigg)
e^{\imath l_i \sqrt{\varepsilon_i k^2-q^2}} &
\bigg(1-\sqrt{\varepsilon_i k^2-q^2\over\varepsilon_{i+1} k^2-q^2}\bigg)
e^{-\imath l_i \sqrt{\varepsilon_i k^2-q^2}} \cr
\bigg(1-\sqrt{\varepsilon_i k^2-q^2\over\varepsilon_{i+1} k^2-q^2}\bigg)
e^{\imath l_i \sqrt{\varepsilon_i k^2-q^2}}&
\bigg(1+\sqrt{\varepsilon_i k^2-q^2\over\varepsilon_{i+1} k^2-q^2}\bigg)
e^{-\imath l_i \sqrt{\varepsilon_i k^2-q^2}}
}\right)
\end{equation}
The transfer matrix connecting the field amplitudes
 in the $i$-th and $(i+1)$-th layer evidently has the form:
\begin{equation}
\left(\matrix{A_{i+n}\cr B_{i+n}}\right)=U_{i+n}^{-1}\hskip1mm
 T_{i+n-1\rightarrow i+n}\hskip1mm
T_{i+n-2\rightarrow i+n-1}\hskip1mm...
T_{i\rightarrow i+1}\hskip1mm
U_i \left(\matrix{A_i\cr B_i}\right)
\end{equation}
 Turn now to the problem of Bragg mirror. We assume that the first period of this
mirror consists of two layers: the first one, with the thickness $l_1$ and
permittivity $\varepsilon_1$, located at a distance of $l_0$ from the origin of the coordinate
system, and the second one with the thickness $l_2$ and permittivity $\varepsilon_2$. The
transfer matrix corresponding to one period of such a Bragg mirror (we denote it
by $T$) will have the form (ct. with (16)):
\begin{equation}
T=T_{2\rightarrow 3}\hskip1mm T_{1\rightarrow 2}
=\left(\matrix{Q & P\cr P^\ast & Q^\ast}\right)
\end{equation}
where
\begin{equation}
Q={1\over 4}\bigg\{\bigg(1+{1\over r}\bigg)\bigg(1+r\bigg)\hskip1mm
e^{\imath(\phi_1+\phi_2)}+
\bigg(1-{1\over r}\bigg)\bigg(1-r\bigg)\hskip1mm
e^{\imath(\phi_1-\phi_2)}\bigg\}
\end{equation}
\begin{equation}
P={1\over 4}\bigg\{\bigg(1+{1\over r}\bigg)\bigg(1-r\bigg)\hskip1mm
e^{\imath(\phi_2-\phi_1)}+
\bigg(1-{1\over r}\bigg)\bigg(1+r\bigg)\hskip1mm
e^{-\imath(\phi_1+\phi_2)}\bigg\}
\end{equation}
and
\begin{equation}
r=\sqrt{\varepsilon_1 k^2-q^2\over\varepsilon_2 k^2-q^2},
\hskip6mm
\phi_i=l_i\hskip1mm \sqrt{\varepsilon_i k^2-q^2}, \hskip2mm i=1,2
\end{equation}
 Let us assume that the mirror comprises $n$ periods, and the media above and
below the Bragg mirror are characterized by the permittivities $\varepsilon_1$
  and $\varepsilon_0$,
respectively. Then the transfer matrix connecting the field amplitudes below and
above the mirror has the form
\begin{equation}
\left(\matrix{A_{2n+1}\cr B_{2n+1}}\right)=U_{2n+1}^{-1}\hskip1mm
T^n T_{0\rightarrow 1}\left(\matrix{A_{0}\cr B_{0}}\right)
\end{equation}
The matrix $T$ entering (22) is determined by the expression
\begin{equation}
T_{0\rightarrow 1}={1\over 2}\left(\matrix{
\bigg(1+\sqrt{\varepsilon_0 k^2-q^2\over\varepsilon_{1} k^2-q^2}\bigg)
e^{\imath l_0 \sqrt{\varepsilon_0 k^2-q^2}} &
\bigg(1-\sqrt{\varepsilon_0 k^2-q^2\over\varepsilon_{1} k^2-q^2}\bigg)
e^{-\imath l_0 \sqrt{\varepsilon_0 k^2-q^2}} \cr
\bigg(1-\sqrt{\varepsilon_0 k^2-q^2\over\varepsilon_{1} k^2-q^2}\bigg)
e^{\imath l_0 \sqrt{\varepsilon_0 k^2-q^2}}&
\bigg(1+\sqrt{\varepsilon_0 k^2-q^2\over\varepsilon_{1} k^2-q^2}\bigg)
e^{-\imath l_0 \sqrt{\varepsilon_0 k^2-q^2}}
}\right)
\end{equation}
 The sought reflectivity can be found from Eq. (22) where $A_0=1$ and $B_{2n+1}=0$. From
 the system of equations thus obtained, the reflectivity can be found as:
\begin{equation}
\ae={B_0\over A_0}=B_0
\end{equation}
 It is seen from Eq. (22) to calculate reflectivity of the Bragg mirror
consisting of $n$ periods, one has to calculate $n$-th power of the matrix $T$. To do
this, let us find eigen numbers   $\lambda_\pm$ and eigen vectors $e^\pm$ of this matrix and define
the matrix $E$ so that its columns are the above eigen vectors:
\begin{equation}
E=\left(\matrix{e^+_1 & e^-_1\cr e^+_2 & e^-_2}\right)
\end{equation}
Then, if $I=E^{-1}$, we have
\begin{equation}
T^n=E\left(\matrix{\lambda_+^n & 0\cr  & \lambda_-^n}\right)I
\end{equation}
The calculations show that
\begin{equation}
\lambda_\pm=\hbox{Re}\hskip1mm Q\pm\sqrt{\hbox{Re}^2 Q-1}
\end{equation}
\begin{equation}
E=\left(\matrix{1 & 1\cr {\lambda_+-Q\over P} &{\lambda_--Q\over P} }\right)
\end{equation}
\begin{equation}
I={1\over \lambda_--\lambda_+}
\left(\matrix{\lambda_--Q & -P \cr Q-\lambda_+ &P }\right)
\end{equation}
Note that matrix (23)  can be represented as
\begin{equation}
T_{0\rightarrow 1}= {1\over 2}\left(\matrix{
\bigg(1+\sqrt{\varepsilon_0 k^2-q^2\over\varepsilon_{1} k^2-q^2}\bigg)
 &
\bigg(1-\sqrt{\varepsilon_0 k^2-q^2\over\varepsilon_{1} k^2-q^2}\bigg)
\cr
\bigg(1-\sqrt{\varepsilon_0 k^2-q^2\over\varepsilon_{1} k^2-q^2}\bigg)
&
\bigg(1+\sqrt{\varepsilon_0 k^2-q^2\over\varepsilon_{1} k^2-q^2}\bigg)
}\right)\hskip1mm U_1\equiv W U_1
\end{equation}
 Then, the Bragg mirror reflectivity can be found from (22):
\begin{equation}
\ae=-e^{2\imath l_0\sqrt{\varepsilon_0 k^2-q^2}}\hskip2mm
{T^n_{21}W_{11}+T^n_{22}W_{21}\over T^n_{21}W_{12}+T^n_{22}W_{22} }
\end{equation}
 Using Eq. (26), this formula can be rewritten as
\begin{equation}
\ae=\ae_n=-e^{2\imath l_0\sqrt{\varepsilon_0 k^2-q^2}}\hskip2mm
{E_{22}I_{21}W_{11}+E_{22}I_{22}W_{21}+\xi^n[E_{21}I_{11}W_{11}+E_{21}I_{12}W_{21}]\over
E_{22}I_{21}W_{12}+E_{22}I_{22}W_{22}+\xi^n[E_{21}I_{11}W_{12}+E_{21}I_{12}W_{22}]
}
\end{equation}
where
\begin{equation}
\xi={\lambda_+\over \lambda_-}
\end{equation}
 while the elements of the matrices  $E$, $I$ and $W$ entering this equation are determined
by Eqs. (28)-(30). Thus the dependence of the reflectivity on the number of
periods of the mirror is given explicitly by Eq. (32). The limiting value of
this coefficient (at   $n\rightarrow\infty$) exists only when $|\xi|\ne 1$.
  One can see from (27) that if $|$Re $Q|<1$,
   then $|\lambda_\pm|=1$ and, hence, $|\xi|=1$.
    Thus the condition of total reflection for
the Bragg mirror can be written as
\begin{equation}
|\hbox{Re}\hskip1mm Q|>1
\end{equation}
Using Eq. (19), this condition can be rewritten in the form:
\begin{equation}
\bigg|\cos\phi_1\hskip1mm\cos\phi_2-
{1\over 2}\bigg(r+{1\over r}\bigg)\sin\phi_1\hskip1mm\sin\phi_2\bigg|>1
\end{equation}
 It is seen that this condition is satisfied, e.g., when $\phi_1=\phi_2=\pi/2$
 and $\varepsilon_1\ne\varepsilon_2$, i.e., when
the mirror consists of quarter-wave layers. In this case, Re $Q<-1$ and, therefore,
$|\xi|<1$. Then, from Eq. (32), we have the following expression for the limiting
reflectivity:
 \begin{equation}
 \ae_\infty=-e^{2\imath l_0\sqrt{\varepsilon_0 k^2-q^2}}\hskip2mm
{I_{21}W_{11}+I_{22}W_{21}\over I_{21}W_{12}+I_{22}W_{22} }
\end{equation}
 The matrix elements entering this equation are given by Eqs. (29) and (30). It
can be shown that the module of this expression always equals unity, i.e.,
\begin{equation}
{I_{21}W_{11}+I_{22}W_{21}\over I_{21}W_{12}+I_{22}W_{22}
}=e^{\imath \theta}, \hskip3mm \hbox{with} \hskip2mm\theta
\hskip2mm
 \hbox{-- real-valued},
\end{equation}
Thus, when reflecting from the infinite Bragg mirror, the wave experiences only
variation of its phase, while its amplitude remains intact. Using Eqs. (7),
(36), and (37), we can write the dispersion law for the {\it empty} ideal Bragg
waveguide in the form similar to (7):
\begin{equation}
2l_0\sqrt{\varepsilon_0 k^2-q^2} +\theta(\omega,q) =\pi m
\end{equation}

Where $m$ is integer and $\theta(\omega,q)$ is determined by Eq. (37).

\section{Propagation of optical pulse in the Bragg waveguid}

 In this section, we will analyze the possibility of reduction of group velocity
of an optical pulse propagating in the Bragg waveguide described above. We will
characterize the optical pulse by the carrier frequency $\Omega$ and spectral density
$\rho(\omega-\Omega)$. The function $\rho(\nu)$ is peaked at $\nu=0$
  and has the width $\Delta\omega$. When such a pulse is
propagating through the waveguide, the field strength as a function of the
coordinate $x$ abd time $t$ can be presented in the form:
   \begin{equation}
   E(x,t)=\int \rho(\omega-\Omega)\exp\imath[\omega t+q(\omega) x] \hskip1mmd\omega
   \end{equation}
 where function $q(\omega)$ provides the dispersion law of the waveguide. In conformity
with the above conclusions, to obtain a high group refractive index $n_g$, the cutoff
frequency $\bar\omega$ of the waveguide should be chosen as close as possibly to the pulse
carrier frequency $\Omega$. At the same time, the pulse spectrum should not come out of
the region of waveguide transmission - otherwise, the pulse will be strongly
distorted and attenuated. Therefore, the cutoff frequency $\bar\omega$, the carrier
frequency $\Omega$, and the spectral width of the pulse $\Delta\omega$
  should meet the following inequality:
\begin{equation}
\Omega-\bar\omega\ge\Delta\omega
\end{equation}

Assuming here that $\Omega-\bar\omega=\Delta\omega$, and using Eq. (12), we obtain for the greatest group
refractive index the following expression:
\begin{equation}
n_{gmax}={G\hskip1mm c\over 2\sqrt{\Delta\omega}}
\end{equation}
 When the characteristics $\Omega$ and $\Delta\omega$ of the optical pulse are known, then parameters
of the Bragg waveguide with the cutoff frequency $\bar\omega$, satisfying condition (40),
can be found in a standard way. Let, e.g., we want to creat a TiO$_2$/SiO$_2$-based
Bragg waveguide. In this case, the values of the material constants are:
$\varepsilon_1= 6.76$ (TiO$_2$, the first layer),
$\varepsilon_2=1.96$ (SiO$_2$, the second layer)
$\varepsilon_0=1.96$ (SiO$_2$, the material of the waveguide itself).

Let us choose $\bar\omega$ corresponding to condition (40). Now we will construct the Bragg
mirrors from the following quarter-wave layers
 $l_i=\pi/(2\bar k\sqrt{\varepsilon_i}), i=1,2$.
 Here $\bar k\equiv\bar\omega/c$. Figure 2 (curve 1)
shows frequency dependence of the phase shift $\theta(\omega,0)$ plotted using Eq. (37)
for this case. The same figure shows the reflectivity spectrum for a finite
Bragg mirror at $n$ = 12 (curve 2) with the above parameters of the layers. As is
seen from Fig. 2, the frequency $\bar\omega$ corresponds to the center of the reflection
window, and $\theta(\bar\omega,0)=0$. Thus, if we choose the half-thickness of the waveguide equal
to a quarter of the wavelength (i.e., so that $2l_0\bar k\sqrt\varepsilon_0=\pi$),
 the Eq. (38) is satisfied at $m=1$, $\omega=\bar\omega$ and $q=0$,
  and, therefore, $\bar\omega$ will be the cutoff frequency of the principal
mode of the waveguide described above. In accordance with the results of Sect.
2, the dispersion law of the waveguide in the vicinity of the cutoff frequency
$\bar\omega$ will have the form of Eq. (11), where the coefficient $G$ can be calculated using
Eq. (11) with the derivatives determines from Eq. (36) (these derivatives can be
easily computed numerically). After that, the maximum group refractive index can
be evaluated using Eq. (41). The calculations performed for the above Bragg
waveguide show that $n_g$ can be ~ 1000 and higher for the spectral width of the
optical pulse of the order of  $\Delta\omega\sim  10^9$ s$^{-1}$.

Of great importance is the question about distortion of the pulse in the process
of its propagation due to strong group velocity dispersion in the vicinity of
the cutoff frequency $\bar\omega$. This question can be answered by direct calculations of
propagation of the pulse with a given spectrum   $\rho(\omega-\Omega)$ using Eq. (39) with the
dispersion law $q(\omega)$  calculated by Eq. (11). The results of these calculations
are presented in Fig. 3, which shows propagation of an optical pulse, $10^{-9}$ s
long, in the Bragg waveguide described above (fig. 3a) for the case when
spectral position of the pulse with respect to the cutoff frequency corresponds
to equality in Eq. (40) (fig.3b). In this case, the group refractive index
attains its maximum value determined by Eq. (41) and equals $n_g=1400$. It is seen from
Fig. 3a that even in this limiting case (when one could expect a considerable
distortion of the pulse upon its propagation) it proves possible to obtain
relative delay (the delay expressed in the units of pulsewidth) of about 10 in
the waveguide ~ 2 mm in length with the two-fold spatial broadening of the
pulse. Similar calculations performed for the case of stronger satisfaction of
condition (40) (when the pulse spectrum is shifted from the cutoff frequency by
more than its spectral width) show that it is possible to obtain the group
refractive index $n_g \sim 500$ and the relative pulse delay $\sim$ 20 - 30 with
acceptable distortion of the pulse shape.

\section{The effects of leakage in a nonideal Bragg waveguid}

Bragg mirrors of real waveguides contain a finite number of periods, and the
reflectivity module of such mirrors is always smaller than unity. This gives
rise to attenuation of the waves propagating in the real waveguide. From the
mathematical point of view, finiteness of the real Bragg mirrors is revealed in
the fact that the dispersion law of the real waveguide differs from that for the
ideal one by appearance of a complex-valued correction $\Delta q(\omega)$
($q(\omega)\rightarrow q(\omega)+\Delta q(\omega)$). The equation that
defines the dispersion law will have, as before, the form of (5), but now it
will be impossible to reduce it to real-valued equation similar to (7) and (38).
Assuming that reflection from the real Bragg mirror is still sufficiently high,
we can obtain the complex-valued dispersion law of the real waveguide using the
perturbation theory in the following way. The deviation of the reflectivity of
the finite Bragg mirror from that of the infinite mirror
 $\Delta\ae=\ae(n<\infty)-\ae(n=\infty)=\ae-\ae_\infty$
 is assumed to be
small. This means that the parameter $\mu\equiv\xi^n$  (see Eq. (32)) is small also. Then, by
differentiating (5) in terms of $\mu$ and $q$ (in this case, the reflectivity in (5)
should be taken in the form of (32)), we obtain:

\begin{equation}
 {\partial\ae^2\over\partial q}\Delta q+
{\partial\ae^2\over\partial \mu}\Delta\mu=
 2\ae\bigg( {\partial\ae\over\partial q}\Delta q+
{\partial\ae\over\partial \mu}\Delta\mu\bigg)=0,
\end{equation}
 whence, for the correction $\Delta q(\omega)$
connected with the finiteness of the Bragg mirrors, we have the expression:
\begin{equation}
\Delta q(\omega)=-{\partial\ae/\partial\mu\over\partial\ae/\partial
q}\Delta\mu={\ae_\infty(\omega,q)-\ae_n(\omega,q)\over \partial\ae_\infty/\partial q}\bigg|_{q=q(\omega)}
\end{equation}
where $q(\omega)$ is the dispersion law of the ideal Bragg waveguide. The derivatives and
increments entering this equation can be easily calculated numerically using Eq.
(32). Now, the quantity $\Delta q(\omega)$ appears to be complex, which corresponds to appearance
of attenuation in the Bragg waveguide. As an example, we have calculated
attenuation of an optical pulse in the waveguide described in the previous
section for the case of 14-period Bragg mirrors. The results are presented in
Fig. 3a, which shows that, in this case, the decrease of the pulse amplitude due
to its broadening and due to its leakage prove to be comparable.

\section{Bragg waveguide with a quantum well inside}

 Let now a thin layer with a high value of permittivity be placed in the $xy$-plane
in the middle of the waveguide. Then the permittivity in the region between the
Bragg mirrors will depend on $z$ as:
\begin{equation}
\varepsilon(z)=\varepsilon_0+\eta\delta(z)
\end{equation}
Under such a layer, we will understand a quantum well, which can be grown inside
the Bragg waveguide using MBE technique. In this case, the frequency dependence
of $\eta$ will show the resonance behavior:
 \begin{equation}
 \eta=\eta(\omega)=a \hskip1mm{\omega_1\over \omega_0-\omega}
 \end{equation}
where  $\omega_1$ characterized the strength of interaction of the exciton localized in
the quantum well with the electromagnetic field, while $a$ is the parameter with
the dimension of length, which can be identified, to a certain extent, with the
well width. The boundary conditions at $z = 0$, in this case, have the form \cite{Koz}:
\begin{equation}
E_y(z=+0)=E_y(z=-0)\hskip10mm
{\partial E_y\over \partial z}\bigg|_{z=+0}-
{\partial E_y\over \partial z}\bigg|_{z=-0}=
-k^2\eta E_y(z=0)
\end{equation}
It follows from these conditions that the field amplitudes above
($A_0$ and $B_0$)
 and below ($\tilde A_0$ and $\tilde B_0$)
the quantum well should obey the relationships (see Sect. 2, Eqs. (2) - (5) and
explanations in the text):
\begin{equation}
\left(\matrix{\tilde{A}_0\cr \tilde{B}_0}\right)=
\left(\matrix{ 1-\imath g & -\imath g  \cr \imath g & 1+\imath g}\right)
\left(\matrix{{A_0}\cr {B_0}}\right)
\end{equation}
  where
\begin{equation}
g\equiv{k^2\eta\over 2 \sqrt{\varepsilon_0 k^2-q^2}}
\end{equation}
Presence of the Bragg mirrors leads, as before, to Eqs. (2) and (4). >From the
condition of nonzero solution of the system of equations (2), (4), and (47) we
can obtain the following equation for the wave dispersion in the studied
composite structure (waveguide + quantum well):
\begin{equation}
\imath g (1+\ae)^2=1-\ae^2
\end{equation}
This equation always has a solution independent of $g$
\begin{equation}
\ae=-1,
\end{equation}
which shows that the presence of the quantum well does not affect the dispersion
laws of odd modes of the {\it empty} waveguide, because these modes have a node in the
plane of the wuantum well. We are interested in the remaining solutions
satisfying the equation:
\begin{equation}
\imath g ={1-\ae\over 1+\ae}
\end{equation}
 If we deal with an ideal Bragg waveguide, then the module of the
 reflectivity $\ae$ equals unity and, hence,
\begin{equation}
 \ae=e^{\imath\phi}\hskip10mm\hbox{with}\hskip3mm \phi=
 \theta(\omega,q)+2 l_0\sqrt{\varepsilon_0 k^2-q^2}+\pi
\end{equation}
 Then, using Eq. (51), we can obtain the following real-valued dispersion relation:
 \begin{equation}
 g=-\hbox{tg}\bigg({\phi\over 2}\bigg)
 \end{equation}
 Denoting resonance frequency of the quantum well by $\omega_0$,
 we can rewrite Eq. (53) as
\begin{equation}
\theta(\omega,q)+ 2 l_0\sqrt{\varepsilon_0 k^2-q^2}+
2\hskip1mm\hbox{arctg}\bigg({k^2 a \omega_1\over 2 \sqrt{\varepsilon_0 k^2
-q^2}(\omega-\omega_0)}\bigg)=(2m-1)\pi,
\end{equation}
 where $m$ is an integer.
 This equation differs from (38) by the third term of the left-hand side, which
is related to the presence of the quantum well in the waveguide. This term gives
rise to appearance of the isolated dispersion branch in the vicinity of the
quantum well resonance frequency. This occurs even when the frequency $\omega_0$ lies
below the cutoff frequency $\bar\omega$ of the empty waveguide. In this case, in the vicinity
of the quantum well resonance frequency there arises a transmission window. The
upper bound of the window coincides with $\omega_0$ and its spectral width is
proportional to the strength of interaction between the quantum well and the
electromagnetic field $a\omega_1$, with the proportionality factor increasing with
approaching the quantum well resonance frequency to the cutoff. The wave
dispersion law in this transmission window has the form of (11) with a fairly
steep dispersion. The relevant calculations of the group refractive index (under
conditions similar to (40)) show that its value may exceed that of the empty
waveguide and may reach $\sim$ 10000 for the pulse with a spectral width of $\delta\omega\sim
10^9$ s$^{-1}$.

 Note, in conclusion, that the possibility of experimental realization of
the above method of the light slowdown in the composite system
(waveguide + quantum well) depends on the value of irreversible
absorption in the quantum well. Evaluation of this absorption from
the measurments of the amplitude and spectral width of the
absorption line can be hampered since such measurements often show
that the spectral width of the excitonic resonance is, to a
considerable extent, of radiative nature.

\begin{figure}
\epsfxsize=400pt
\epsffile{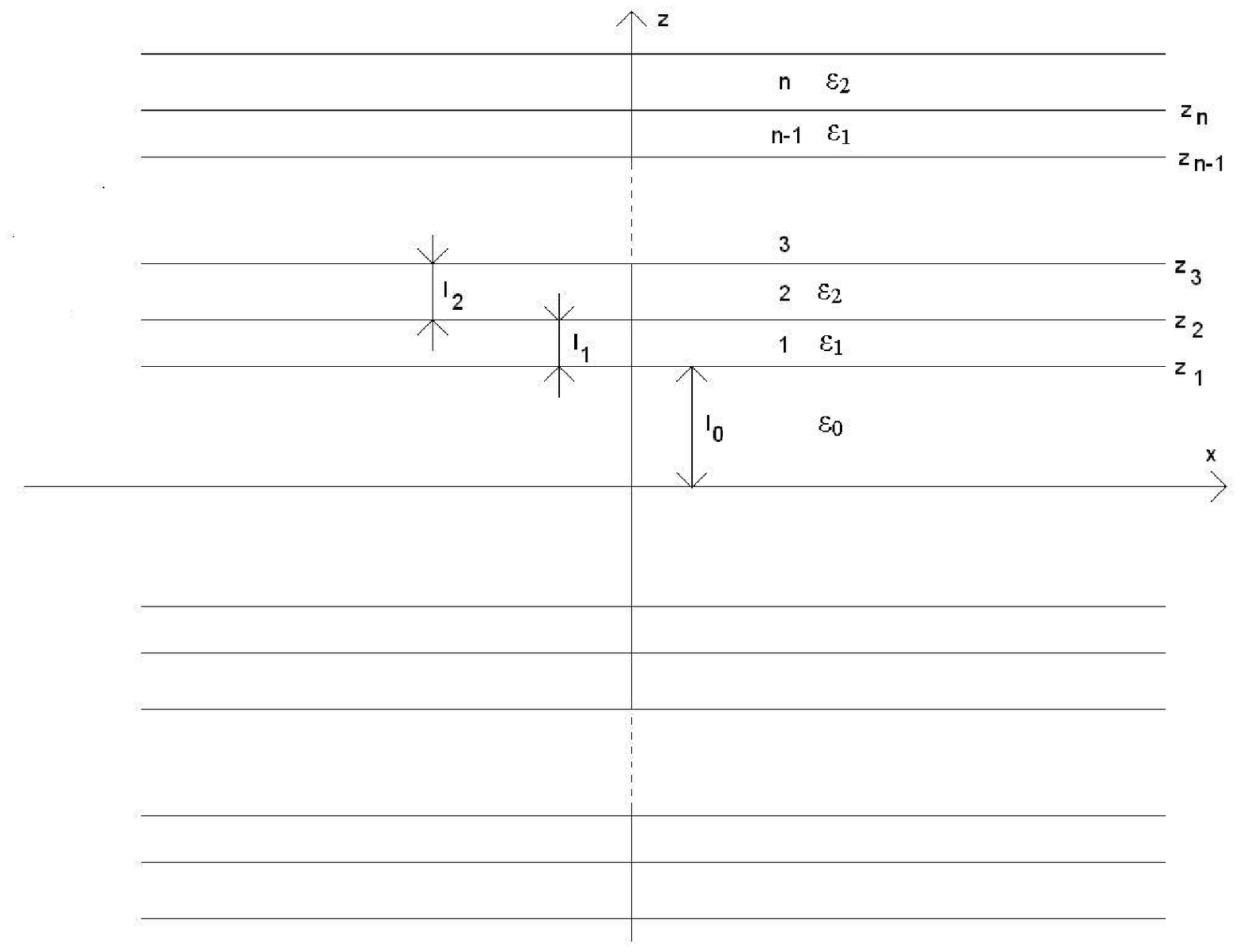}
\caption{}
\end{figure}

\begin{figure}
\epsfxsize=400pt
\epsffile{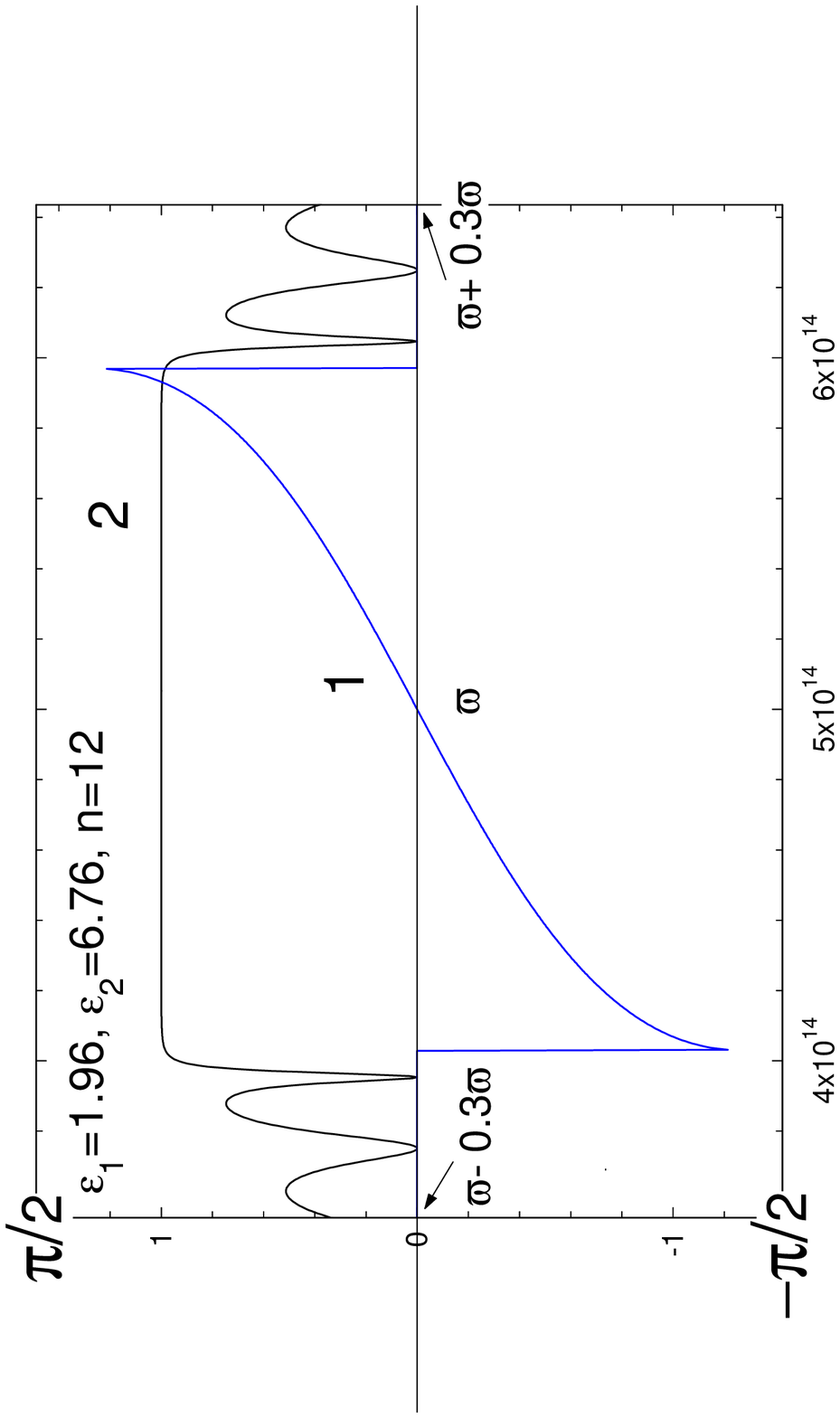}
\caption{}
\end{figure}

\begin{figure}
\epsfxsize=400pt
\epsffile{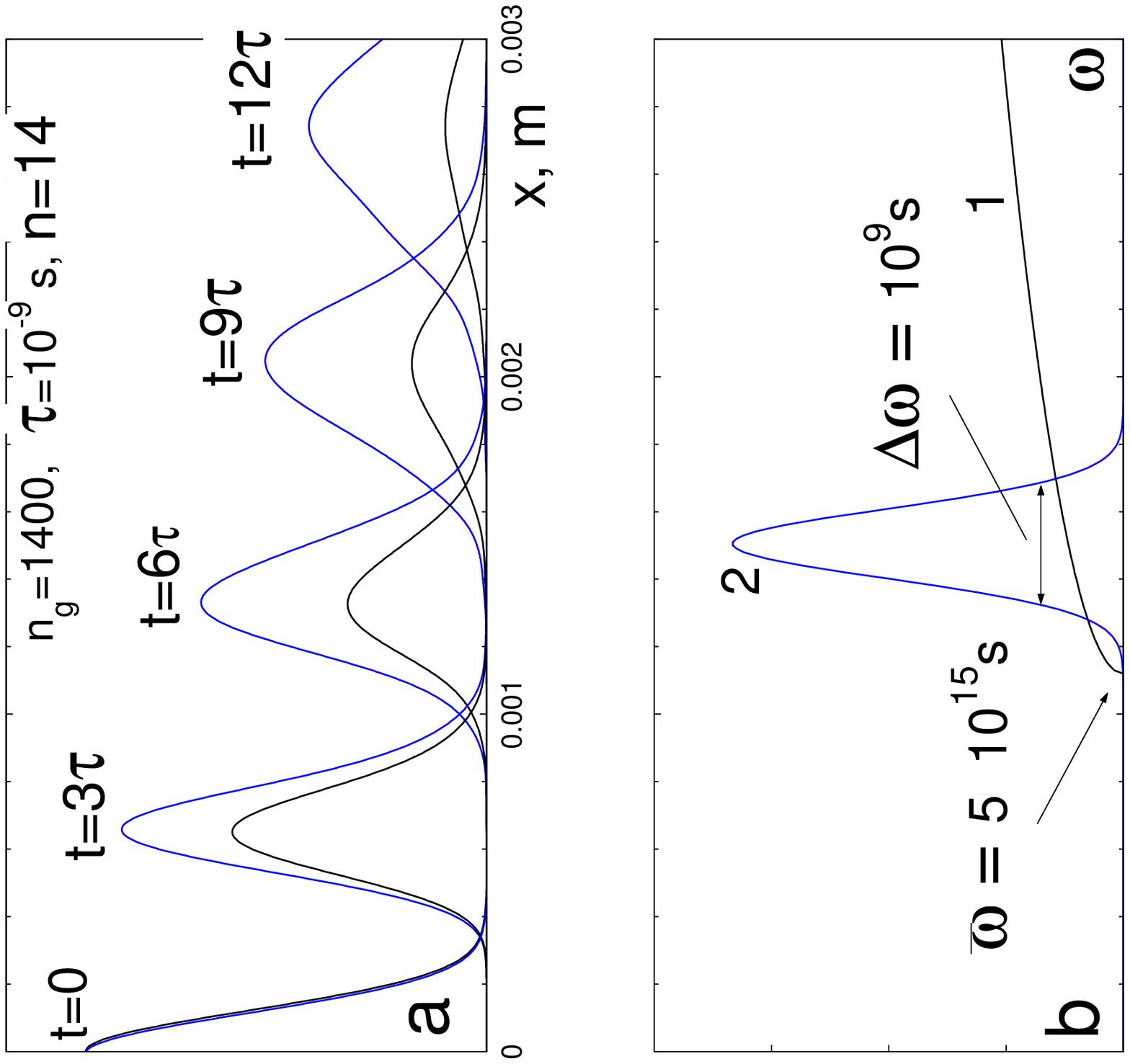}
\caption{}
\end{figure}
\end{document}